\documentclass[aps,notitlepage,a4paper,superscriptaddress,11pt]{article}

\usepackage[utf8]{inputenc}
\usepackage[dvips]{color}
\usepackage[toc]{appendix}
\usepackage[hang,flushmargin]{footmisc} 
\usepackage{titlesec,titletoc,ntheorem-hyper,authblk,abstract,latexsym,microtype,booktabs,amsmath,cleveref,fancyhdr,wrapfig,array,amsfonts, amssymb,geometry,appendix,cite}

\newcommand{\be}{\begin{equation}}
\newcommand{\ee}{\end{equation}}
\newcommand{\bea}{\begin{eqnarray}}
\newcommand{\eea}{\end{eqnarray}}
\newcommand\blfootnote[1]{%
  \begingroup
  \renewcommand\thefootnote{}\footnote{#1}%
  \addtocounter{footnote}{-1}%
  \endgroup
}

\usepackage{graphicx,lipsum}
\usepackage{caption}
\usepackage{subfigure}
\usepackage{amsmath}
\usepackage{amsfonts}
\usepackage{amssymb}
\usepackage{stackengine}
\usepackage{mathtools}
\numberwithin{equation}{section}
\usepackage{titlesec}
\usepackage{sectsty}
\sectionfont{\centering}

\numberwithin{subcase}{case}
\usepackage{framed}
\allowdisplaybreaks
\usepackage[nottoc]{tocbibind}
\usepackage[free-standing-units]{siunitx}
\usepackage{helvet}
\usepackage{url}
\usepackage{titletoc}
\usepackage{blindtext}
\usepackage[gen]{eurosym}

\title{Generalized Uncertainty principle and  momentum-dependent effective mass Sch\"{r}odinger equation}

\author[1]{Bijan Bagchi}
\author[2]{Rahul Ghosh}
\author[3]{Partha Goswami}
\affil[1, 2]{Department of Physics, Shiv Nadar University, Dadri, UP 201314, India}
\affil[3]{Department of Physics, Deshbandhu College, Kalkaji, New Delhi 110019, India}

\begin{document}
	\maketitle	
	
\begin{abstract}

We show in this paper that the basic representations of position and momentum in a quantum mechanical system, that are guided by a generalized uncertainty principle and lead to a corresponding one-parameter eigenvalue problem, can be interpreted in terms of an extended Sch\"{r}odinger equation embodying momentum-dependent mass. Some simple consequences are pointed out.
\end{abstract}

\blfootnote{
    E-mails: {bbagchi123@gmail.com}, {rg928@snu.edu.in}, physicsgoswami@gmail.com}


\maketitle

\maketitle

\section{\label{sec:level1}Introduction}
\maketitle

The general theory of relativity (GTR), whose character is completely classical, and quantum mechanics, which accounts for the properties of atomic and sub-atomic particles but excludes the effects of gravity, are two definitive pillars of twentieth century physics. To unify them requires tuning to the Planck scale. In such a situation, for a particle of mass $m$, the size of the radius of curvature of the spacetime is roughly of the same order as its Compton wavelength. It is simple to derive the connecting relation which reads 

\begin{flalign}
        m \approx m_P\sqrt{\pi}
\end{flalign}
where $ m_P $ denotes the Planck mass. 

Furthermore, as is well-known, if the linear dimension of the mass $m$ is less than the corresponding Schwarzschild radius, the mass mimics a black hole. It follows that the global length measurement uncertainty $\delta X$ must exceed the Schwarzschild radius favoring the inequality

\begin{flalign}
\Delta X \geq \kappa \frac{Gmc}{c^3}
\end{flalign}

where the quantity $ \kappa $ is a dimensionless parameter of order unity. The exact value of  $ \kappa $ depends on the choice of a specific model. 

A stronger inequality has the form $ \Delta X \geq G\gamma \frac{\Delta P}{c^3} $ which can be recast to 

\begin{flalign}
\Delta X \geq \gamma l_P^2 \frac{\Delta P}{\hbar}
\end{flalign}
where $ \Delta P $ stands for the complementary linear momentum uncertainty and $ l_p $ is the Planck length. Combining with the standard Heisenberg's uncertainty principle (HUP), which states that for certain pairs of complementary variables, such as the position and momentum, one cannot have an exact knowledge of both the variables simultaneously, one has the following form of the generalized uncertainty principle (GUP) 

\begin{flalign}
\Delta X \Delta P \geq \frac{\hbar}{2}(1+\tau {\Delta P}^2) \label{DxDp} \
\end{flalign} 
in one dimension where the deformation parameter $\tau = \frac{2\gamma l_P^2}{\hbar^2} > 0$. In arriving at $(1.4)$ a modified form of the quantum condition is used namely,

\begin{flalign}
[X, P] = \hbar (1 + \tau P^2) \label{x,p} 
\end{flalign}
to lowest order in $ \tau $.  

In the framework of GUP a minimum position uncertainty is implied which is of the order of the Planck length. However there is no minimum momentum uncertainty. Note that the conventional Heisenberg's uncertainty principle is recovered from $(1.4)$ when the parameter $ \tau $ vanishes. $ \tau $ is a dimensional quantity being same as that of the inverse squared momentum. For an optical fibre, a typical value of $ \tau $ is $ 10^{56}(\frac{s^2}{Kg^2m^2}) $.

 $ GUP $ which is a modification of the Heisenberg's uncertainty principle, in that it takes into account plausible corrections to the latter, precludes any localization in space as is implicit in the case in the Heisenberg's relation. There are many derivations of GUP existing in the literature. In fact, a considerable amount of work has been devoted towards justifying the viability of having different variants of GUP in problems of minimal length in quantum gravity and string theory (see, for example, \cite{Taw, Mag, gross, amati, hossenfelder, lake, rovelli, das, dey}). In the former case the existence of a minimal length is imperative due to an effective minimal uncertainty in position while in the latter, because of the existence of a characteristic length of the strings, it is impossible to improve upon the spatial resolution below a certain point. 
 
In  a general setting the following result \cite{bagchi1, kempf} was obtained for the commutator $[X,P]$ undergoing a $q$-deformation

\begin{flalign}
[X, P] = i \hbar \left [1 +\frac{i\hbar}{4} (q^2 -1) (\frac{X^2}{a^2}+\frac{P^2}{b^2}) \right ] \label{gen x,p} \
\end{flalign}
where $a$ and $b$ are some real constants and the constraint $ab = \frac{1}{4} (q^2 +1)$ has been taken into account. Adopting for the deformation parameter $q$ the exponential form $q=e^{2\tau b^2}$, the limit $b \rightarrow 0$ yields $(1.6)$. 

As noted in \cite{kempf}, the following explicit representations for $ X $ and $ P $

\begin{flalign}
X = i \hbar(1+\tau p^2)\partial_p ,  \quad P=p  \label{X, P rep}
\end{flalign}
solve $(1.5)$, where $ p $ along with $x$ fulfill the HUP

\begin{flalign}
[x, p] = i\hbar
\end{flalign}

Let us focus on the simplest case of the harmonic oscillator Hamiltonian which  reads (setting $m = \hbar = 1 $)
\begin{flalign}
H=\frac{1}{2}[P^2 + \omega^2 X^2] \label{H_LHO}
\end{flalign}
Inserting the representations $(1.7)$ in $(1.9)$ produces the following stationary state Schr\"{o}dinger equation 
\begin{equation}
 \phi''(p) + \frac{2\tau p}{1+\tau p^2}\phi'(p)
= \frac{1}{(1+\tau p^2)^2}[\mu^2 p^2 - \lambda^2] \phi(p)  \label{HO_SE}
 \end{equation} 
where $\phi (p)$ is the wave function expressed in terms of $p$ and the primes denote derivatives with respect to $p$. In terms of the energy $E$ the quantities $\lambda $ and $\mu $ stand for 

\begin{flalign}
\lambda = \frac{2E}{\omega^2}, \quad \mu^2 = \frac{1}{\omega^2}
\end{flalign}
It is needless to say that the standard equation for the harmonic oscillator is recovered when $\tau \rightarrow 0$. Equation $(1.10)$ is similar to the one derived by Conti \cite{conti1} in the context of non-paraxial setting \cite{conti2, miri} by assuming a specific form of the nonlinear wave equation.

In the operator version defined by $X=(\frac{1}{2\omega})^\frac{1}{2} (\eta +\eta^\dagger), \quad P=i(\frac{\omega}{2})^\frac{1}{2} (\eta^\dagger -\eta)$, the harmonic oscillator Hamiltonian $(1.9)$ takes the usual form $H= \omega (\eta^\dagger \eta +\frac{1}{2})$. For the non-Hermitian counterpart Swanson \cite{swn} proposed the following choice  

\begin{flalign}
H^{(\alpha, \beta)} = \omega (\eta^{\dagger} \eta + \frac{1}{2}) +\alpha \eta^2 +\beta \eta^{\dagger}   \label{H_swanson1}
\end{flalign}
where $\omega, \alpha, \beta$ are real while $\bar{\omega} \equiv \sqrt{\omega^2-4\alpha\beta} >0$ and $ \alpha \neq \beta$ \cite{gey}. $H^{(\alpha, \beta)}$ thus translates to 

\begin{flalign}
H^{(\alpha, \beta)}= (\omega -\alpha -\beta) \frac{P^2}{2\omega} + (\omega+\alpha +\beta)\frac{\omega}{2} X^2  +\frac{1}{2}(\omega + \alpha -\beta)XP - \frac{1}{2}(\omega - \alpha +\beta)PX +\frac{\omega}{2}  \label{H_swanson2}
\end{flalign}
It supports the modified energy eigenvalues $E_n = (n + \frac{1}{2}) \bar{\omega}$.
Evidently $(1.13)$ moves over to $(1.9)$ when $\alpha =\beta =0 $.

By using the pair of representations $(7)$, the corresponding time-independent Schr\'{o}dinger eigenvalue equation turns out to be
\begin{flalign}
\omega(\omega +\alpha +\beta)\phi''(p) + \frac{2p}{1+\tau p^2}[\omega(\omega +\alpha +\beta)\tau +(\alpha -\beta)]\phi'(p) 
\nonumber \\ = [\big(\frac{\omega -\alpha-\beta}{\omega}-(\omega +\alpha -\beta)\tau\big)p^2 -(2E+\alpha -\beta)\big)] \frac{\phi(p)}{(1+\tau p^2)^2}    \label{SE_swanson}
\end{flalign}

 In the following our endeavor would be to interpret $(1.10)$ and $(1.14)$ in the framework of a $\tau$-induced momentum-dependent mass (MDM) effective Schr\'{o}dinger equation for it.

\section{\label{sec:level1}The analogue MDM connection}

First, a few words about the coordinate-dependent mass picture in the configuration space that has been studied widely in the literature \cite{vonroos,bagchi2, dut, mus1, mus2, dha, carinena,cruz,ortiz,fer, cunha}. The interest in such systems stem essentially from the physical problems underlying compositionally graded crystals \cite{Gel}, quantum dots \cite{Ser}, liquid crystals \cite{Bar} etc. Likewise, a MDM scenario has also found relevance in certain quantum mechanical problems such as in the quantization of a parity-symmetric Li\'{e}nard type nonlinear oscillator \cite{ruby} and also in classical problems possessing quantum analogs, for example, in branched Hamiltonian systems \cite{bagchi3, bagchi4}.

Many years ago, von Roos \cite{vonroos} formulated a general strategy of writing down an effective-mass kinetic energy operator $\hat{T}$. For instance if the mass function depends on the momentum then $\hat{T}$ would depend on the two- parameter form  (with $\hbar=1)$

\begin{flalign}
\hat{T} = \frac{1}{4}[m^a (p)\hat{x}m^b (p)\hat{x}m^c (p) + m^c (p)\hat{x}m^b (p)\hat{x}m^a (p)] \label{KE_VonR}
\end{flalign}
where $ \hat{x} $ is the coordinate operator, $m(p)$ is a momentum dependent mass and the ambiguity parameters $a,b$ and $c$ are constrained by the equality $ (a +b + c)= -1 $.

For the conventional quantum condition $ [\hat{x}, \hat{p}]=i$, a suitable operator for $ \hat{x} $ is $ [i \partial_p] $ in the momentum space. In this case the above expression of $\hat{T}$ is converted to the following expression for the time-independent Schr\"{o}dinger equation

\begin{flalign}
 -\frac{1}{4} [m^a (p)\frac{d}{dp}m^b (p)\frac{d}{dp}m^c (p) + m^c (p)\frac{d}{dp}m^b (p)\frac{d}{dp}m^a (p) ] \psi +V(p) \psi = \Lambda \psi  \label{SE_MDM1}
\end{flalign}
where $V(p)$ stands for the general choice of a momentum-dependent potential and $\Lambda$ is the energy.

Employing a dimensionless mass function $M(p)$ through enforcing the relation 
\begin{flalign}
 m(p)=m_0 M(p)  
\end{flalign} 
where $2m_0=1$, the identity

\begin{flalign}
 &M^a \frac{d}{dp}M^b \frac{d}{dp}M^b +  M^c \frac{d}{dp}M^b \frac{d}{dp}M^a \nonumber \\ =& \frac{d}{dp}\frac{1}{M(p)}\frac{d}{dp} - (b +1) \frac{M''}{M^2} +2([a(a +b +1) +b +1] \frac{M'^{2}}{M^3}
\end{flalign}
makes a short work of $(2.2)$ casting it to a very tractable form

\begin{flalign}
[-\frac{d}{dp}\frac{1}{M(p)}\frac{d}{dp} + V_{eff}(p)]\phi(p) = \Lambda \phi(p)\label{SE_MDM2}
\end{flalign}
in the presence of an effective potential $V_{eff}(p)$ where 

\begin{flalign}
V_{eff}(p) = V(p) + \frac{1}{2} (b +1)\frac{M''}{M^2}-[a(a +b +1) +b +1]\frac{M'^{2}}{M^3} \label{v_effM1}
\end{flalign}

If we compare $(2.5)$ with the $GUP$ influenced oscillator equation $(1.10)$, we immediately infer that it can be recognized as an analogue momentum-dependent mass equation which transforms similar to $(2.2)$, the mass behaving like a  momentum-dependent quantity. Indeed corresponding to $(1.10)$ the momentum-dependent mass $M(p)$ reads explicitly

\begin{flalign}
M(p)= {(1+\tau p^2)}^{-1}   \label{Massf1}
\end{flalign}
which is independent of the ambiguity parameters $a$, $b$, $c$. Evidently it is singularity free. Such a profile has been studied in \cite{mus2} in a different context. It is clear that $M(p)$ approaches a zero-value asymptotically with respect to $p$. The MDM works in the presence of the $\tau$-dependent effective potential $V_{eff}(p)$ 

\begin{flalign}
   V_{eff}(p)- \Lambda  = \frac{1}{1+\tau p^2}(\mu^2 p^2 - \lambda)  \label{v_eff1}
\end{flalign}
where $\Lambda$ is some constant quantity.

Turning to the more elaborate form $(1.14)$, the corresponding quantities for $M(p)$ and $V_{eff}$ are

\begin{flalign}
M(p)= {(1+\tau p^2)}^{-\bigg[1+\frac{\alpha - \beta}{\omega \tau (\omega+\alpha+\beta)}\bigg]}   \label{Massf2}
\end{flalign}
and 
\begin{flalign}
 V_{eff}(p)- \Lambda = \bigg[ \bigg\{ \frac{\omega-\alpha-\beta}{\omega}-(\omega+\alpha-\beta)\tau \bigg\} p^2 -(2E+\alpha-\beta)\bigg] \nonumber \\ \quad \bigg[\frac{(1+\tau p^2)^{-2+{\big[1+\frac{\alpha - \beta}{\omega \tau (\omega+\alpha+\beta)}\big]}}}{ \omega(\omega+\alpha+\beta)}\bigg]  \label{v_eff2}
\end{flalign}
The trend of the mass profile is similar to the previous one because $\tau$ acts as a huge damping factor.

From the point of view of the intertwining relationship \cite{bagchi2}
\begin{flalign}
AH=H_1A
\end{flalign}
where $ H $ and $ H_1 $ share the same kinetic energy term with $ H_1 $ is defined in terms of an associated potential $ V_{1,eff}(p) $ the following interpretation in the spirit of supersymmetric quantum mechanics \cite{cooper} can be given. If $ A $ annihilates the ground-state wavefunction of $ H $ i.e. $ A\phi_0 =0 $, the eigenvalues of $ H_1 $ obey the relations $ \Lambda_{1,n} = \Lambda_{n+1} $, $ n=0,1,2,.. $.  $ H_1 $ has the corresponding wavefunctions reading $ A\phi_{n+1} $ because of the equality $ H_1(A\phi_{n+1})=AH\phi_{n+1}=\Lambda_{n+1}(A\phi_{n+1}) $ for $ n=0,1,2,... $.

Using now the minimal first-derivative representation of $A$ i.e. $A=\xi(p)\frac{d}{dp} + \theta (p)$, where $\xi(p)$ and $(\theta (p)$ are arbitrary suitable functions, the consistency condition

\begin{flalign}
\xi(p)=\sqrt{1+\tau p^2} 
\end{flalign}
readily follows along with the following forms of $V_{eff}$ and $V_{1,eff}$

\begin{eqnarray}
&& V_{eff}(p) = c+\theta^2-(\xi \theta)',\nonumber \\
&& V_{1, eff}(p)= V_{eff}+2\xi\theta' -\xi'\xi''   \label{fac1}
\end{eqnarray}
where $c$ is an integration constant. Specifically, we obtain for $\kappa(p)$ the zero-energy result

\begin{flalign*}
\kappa(p) = \frac{\eta p}{\sqrt{1+\tau p^2}}
\end{flalign*}
for $\Lambda$ = $c- \eta$.  \\

Some remarks are in order on the feasibility of a generalized quantum condition by taking into account the most general first-order differential form for $\eta$ in the momentum space \cite{bagchi5}

\begin{flalign}
\eta (p) = r(p) \frac{d}{dp} + s(p), \quad r(p), s(p) \in \mathbb{R}
\end{flalign}
It gives the generalized form of the commutator

\begin{flalign}
[\eta, \eta^\dagger ]  = 2rs' - rr^{''}   \label{eta,eta+}
\end{flalign}
in contrast to the standard $[\eta, \eta^\dagger] =1$ considered earlier in this section. In $(2.15)$ the prime denotes derivatives with respect to $p$.

In the light of the new commutator $(2.15)$, the Hamiltonian $H^{(\alpha, \beta)}$ assumes the form

\begin{flalign}
H^{(\alpha, \beta)} \rightarrow \tilde{H}^{(\alpha, \beta)}(p) = -\frac{d}{dp} \tilde{r}^2 (p)\frac{d}{dp} + \tilde{s} (p) \frac{d}{dp} + \tilde{w} (p) \label{H_swanson_dif}
\end{flalign}
where the coefficients $\tilde{r}$, $\tilde{s}$, $\tilde{w}$ stand for the quantities

\begin{eqnarray}
&& \tilde{r} = \sqrt{\tilde{\omega}}r \quad \tilde{\omega} =\omega -\alpha -\beta >0, \label{fac1} \\
&& \tilde{s} = (\alpha - \beta)(2rs - rr'),   \label{fac2}\\
&& \tilde{w} = \omega (s^2 -rs' -r's) +\alpha (rs' +s^2) +\beta (rr''+r'^2 -rs' -2r's +s^2) +\frac{\omega}{2}  \label{fac3} \quad
\end{eqnarray}

Applying the similarity transformation on $\tilde{H}^{(\alpha, \beta)}(p)$ according to

\begin{eqnarray}
&&\tilde{h}^{(\alpha, \beta)} = \rho^{(\alpha, \beta)}\tilde{H}^{(\alpha, \beta)}[\rho^{(\alpha, \beta)}]^{-1}   \label{H_hermitian1} , \\
&&\rho^{(\alpha, \beta)} = e^{-\frac{1}{2} \int^p \frac{\tilde{s}}{\tilde{r}^2}dp}  \label{rho_hermitian}
\end{eqnarray}
produces the compact form

\begin{flalign}
\tilde{h}^{(\alpha, \beta)} = -\frac{d}{dp} \tilde{r}^2 (p)\frac{d}{dp} + \tilde{V}^{(\alpha, \beta)} (p)  \label{H_hermitian2}
\end{flalign}
with $ \tilde{V}^{(\alpha, \beta)} $ being

\begin{flalign}
\tilde{V}^{(\alpha, \beta)} (p) = \frac{\tilde{s}^2}{4\tilde{r}^2} - \frac{1}{2}\tilde{s}' + \tilde{w}
\end{flalign}
Comparison with $(2.5)$ and subsequently to $(1.14)$ reveals the mass function to be dependent on the parameter $ \tau $ if the p-dependent function $r$ is explicitly known. Further knowledge of $r(p)$ and $s(p)$ defines the form of the potential $ \tilde{V}^{(\alpha, \beta)} $.

\section{\label{sec:level1} Summary}

To summarize, we have made in this paper an interesting observation that the eigenvalue problem resulting from the GUP picture can be interpreted in terms of a MDM mechanical system by making appropriate identifications. In this connection we also exploited the intertwining relationships between two partner effective potentials and derived a set of consistency relations.

\section{Acknowledgment}

One of us (BB) thanks Axel Schulze-Halberg for helpful comments.

\end{document}